\documentclass[runningheads]{llncs}
\usepackage{amsmath}
\usepackage{mathtools}
\usepackage{graphicx}
\usepackage{graphics}
\usepackage{svg}
\usepackage{rotating}

\usepackage[colorinlistoftodos]{todonotes}
\usepackage[colorlinks=true, allcolors=blue]{hyperref}
\usepackage{caption}
\DeclareCaptionFont{mysize}{\fontsize{9}{9}\selectfont}
\usepackage[title]{appendix}
\usepackage{amsfonts}
\usepackage{multirow}
\usepackage{booktabs}
\usepackage[ruled,vlined,algo2e]{algorithm2e}

\newcommand{\R}{\mathbb{R}}

\usepackage{acro}
\DeclareAcronym{PET}{
short=PET,
long=Positron Emission Tomography
}
\DeclareAcronym{SE}{
short=SE,
long=squeeze and excitation
}

\def\@fnsymbol#1{\ensuremath{\ifcase#1\or *\or \dagger\or \ddagger\or
   \mathsection\or \mathparagraph\or \|\or **\or \dagger\dagger
   \or \ddagger\ddagger \else\@ctrerr\fi}}
\newcommand{\ssymbol}[1]{^{\@fnsymbol{#1}}}

\newcommand\Tstrut{\rule{0pt}{2.4ex}}         
\newcommand\Bstrut{\rule[-0.9ex]{0pt}{0pt}}   

\begin{document}

\title{Simultaneous Denoising and Motion Estimation for Low-dose Gated PET using a Siamese Adversarial Network with Gate-to-Gate Consistency Learning}
 

\author{Bo Zhou\inst{1} \and Yu-Jung Tsai\inst{2} \and Chi Liu\inst{1,2}}

\institute{Biomedical Engineering, Yale University, New Haven, CT, USA \and Radiology and Biomedical Imaging, Yale University, New Haven, CT, USA}

\authorrunning{B. Zhou, etc} 
\titlerunning{Denoising and motion estimation for low-dose gated PET} 

\maketitle              

\begin{abstract}
Gating is commonly used in PET imaging to reduce respiratory motion blurring and facilitate more sophisticated motion correction methods. In the applications of low dose PET, however, reducing injection dose causes increased noise and reduces signal-to-noise ratio (SNR), subsequently corrupting the motion estimation/correction steps, causing inferior image quality. To tackle these issues, we first propose a Siamese adversarial network (SAN) that can efficiently recover high dose gated image volume from low dose gated image volume. To ensure the appearance consistency between the recovered gated volumes, we then utilize a pre-trained motion estimation network incorporated into SAN that enables the constraint of gate-to-gate (G2G) consistency. With high-quality recovered gated volumes, gate-to-gate motion vectors can be simultaneously outputted from the motion estimation network. Comprehensive evaluations on a low dose gated PET dataset of 29 subjects demonstrate that our method can effectively recover the low dose gated PET volumes, with an average PSNR of 37.16 and SSIM of 0.97, and simultaneously generate robust motion estimation that could benefit subsequent motion corrections.

\end{abstract}
\acresetall

\section{Introduction}
\begin{figure}[htb!]
\centering
\includegraphics[width=0.76\textwidth]{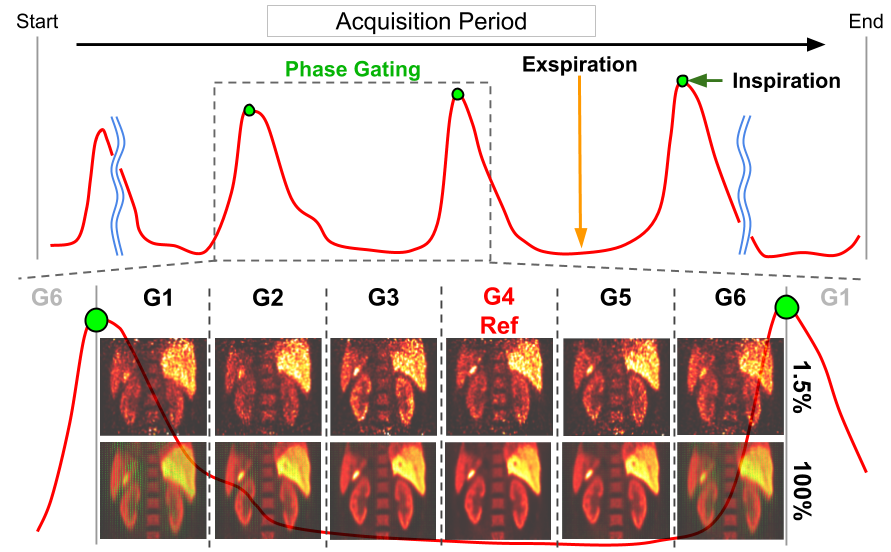}
\caption{Illustration of phase gated PET acquisition with 6 gates for both 100\% full count and 1.5\% count levels. End-expiration gate with the least intra-gate motion (G4) is used as reference gate. Each low dose gated volume needs to be denoised and registered to the reference gated volume. }
\label{fig:illustration}
\end{figure}

PET is a commonly used functional imaging modality. To obtain high quality image, a small amount of radioactive tracer is administered to patient, introducing radiation exposure to both patients and healthcare providers \cite{strauss2006alara}. PET data acquisition typically takes several minutes. During this period, patient’s breathing inevitably introduces blurring in the lung and abdominal regions. Respiratory gating facilitated by external motion monitoring devices are typically used to reduce respiratory-induced motion blurring. However, each gated image is generated by only a fraction of detected events, leading to high image noise in each gate. To address the noise issue, previous works proposed motion correction approaches involving non-rigid image registration among gated images, and utilize the motion vectors to correct motion by using all detected events to reduce image noise \cite{catana2015motion}. In the applications of radiation dose reduction, reduction of injection dose is the first choice but will increase the image noise and result in low signal-to-noise ratio (SNR). In the cases where respiratory gating is performed using low-dose data, the image noise is further increased, potentially causing errors in motion vector estimation, which subsequently affects the final motion correction results, as illustrated in Figure \ref{fig:illustration}. To address this challenge, we aim to simultaneously tackle the image denoising and motion estimation problems in low-dose gated PET data.

Previous works on denoising low-dose PET can be summarized into two categories: conventional post-processing \cite{dutta2013non,maggioni2012nonlocal,mejia2016noise} and deep learning based post-processing \cite{xiang2017deep,wang20183d,lu2019investigation,kaplan2019full}. Conventional post-processing techniques, such as Gaussian filtering, is the standard technique to reduce PET image noise, but has challenge to preserve local structure. More recently, non-local mean filter \cite{dutta2013non} and block-matching 4D filter \cite{maggioni2012nonlocal} have been proposed to denoise low-dose PET while better preserving the structural information. Deep learning based methods, such as deep auto-context CNN \cite{xiang2017deep}, 3D cGAN \cite{wang20183d}, UNet \cite{lu2019investigation}, and GAN \cite{kaplan2019full}, were developed for recovering standard-dose PET from low-dose PET. Compared to conventional methods, these deep learning based methods achieved promising denoising performance on static low-dose PET. However, none of these previous studies addressed denoising and motion estimation in low-dose respiratory gated PET in a unified fashion. 

In this work, we proposed a Siamese adversarial network (SAN) with gate-to-gate consistency learning (G2G) to simultaneously denoise low dose gated volumes and estimate the motion among the gates. We evaluated our method on a challenging low dose gated PET dataset with only $1.5\%$ count level. Our experimental results demonstrated that our proposed method can effectively reduce the noise while preserving the structural information and improve the accuracy of motion estimation.

\section{Problem Formulation}
Assuming a phase gated PET exam generates 6 gates with gate 4 as the reference gate, we denote high-dose PET (HDPET) and low-dose PET (LDPET) gated volumes as $H_n , L_n \in \R^{h \times w \times d}$ with gate index $n \in \{1,2,3,4,5,6\}$ and volume size $h \times w \times d$. The transformation predicted between $\{ L_4,L_n \}$ is expected to be different from the transformation predicted between $\{ H_4,H_n \}$ due to the high noise level of LDPET. Given that the distribution of HDPET is unknown, our goal is to recover $H_n$ from the degraded $L_n$. Previous methods have been trying to solve the inverse problem by finding the generative model $\mathcal{P}_D$ parameterized by $\theta_D$ such that $ \mathcal{P}_D (\sum_{n=1}^{6} L_n ; \theta_D) = \bar{H}_{nmc} \approx H_{nmc} $, where $\bar{H}_{nmc}$ is the non-gated denoised volume with no motion correction (nmc). Since no motion estimation and corresponding motion correction component are considered, degradation in the final image can be expected. Therefore, we aim to tackle these issues by recovering the HDPET from LDPET for each gate and simultaneously estimate the motion field between gates. Specifically, we want to find single gate generative model $\mathcal{P}_D$ such that $\mathcal{P}_D (L_n ; \theta_D) = \bar{H}_n \approx H_n $ where $\bar{H}_n$ is the recovered HDPET for gate $n$. Then, the motion transformation between the reference gate (assume to be gate 4 here) and gate $n$ is estimated by $ \bar{T}_n = \mathcal{P}_R (\mathcal{P}_D (L_4; \theta_D), \mathcal{P}_D (L_n ; \theta_D)  ; \theta_R) \approx T_n $, where $\bar{T}_n$ is the predicted transformation from our motion estimator $\mathcal{P}_R$. In this work, our goal is to obtain the optimal $\mathcal{P}_D$ and $\mathcal{P}_R$ for simultaneous denoising and motion estimation.

\section{Methods}
\begin{figure}[htb!]
\centering
\includegraphics[width=1.00\textwidth]{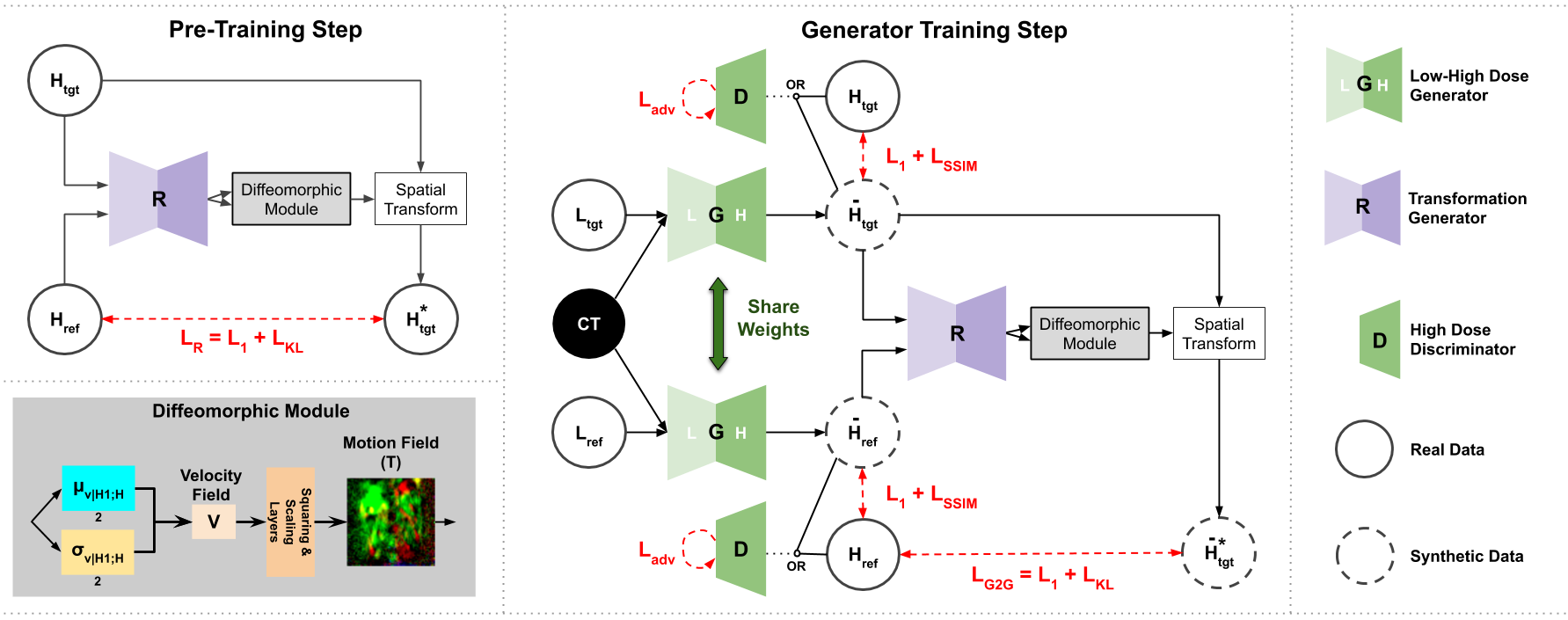}
\caption{Our two-stage training procedure consists of: the pre-training of our motion estimator (R), and Siamese adversarial training of our generator. Two shared weights generators $G$ learn mapping from LDPET to HDPET, which are supervised by a structure recovery loss ($\mathcal{L}_{SR}=\mathcal{L}_{1}+\mathcal{L}_{SSIM}+\mathcal{L}_{adv}$), and a transform consistency loss ($\mathcal{L}_{G2G}=\mathcal{L}_{1}+\mathcal{L}_{KL}$), respectively. Motion estimator $R$ is pre-trained with the ground truth HDPET, and concatenated to the generator for end-to-end optimization. Network architecture details are listed in the supplementary.}
\label{fig:network}
\end{figure}

The overall pipeline of our method is illustrated in Figure \ref{fig:network}. It consists of three major parts: 1) Siamese generative networks with supervision from our structure recovery loss; 2) unsupervised motion estimation network; and 3) gate-to-gate consistency training. The Siamese generator $G$ maps the target gate LDPET ($L_{tgt}$) and the reference gate LDPET ($L_{ref}$) to the HDPET space simultaneously, thus generating denoised HDPET gated volumes. The generator $G$ is first optimized based on the structure recovery loss that measures the dissimilarity between prediction and ground truth, yielding the high quality denoised HDPET volumes. In the meantime, the motion estimation network $R$ is pre-trained using the ground truth HDPET gated volumes $H$, and concatenated to the Siamese generative networks. By replacing the input for $R$ with the synthetic HDPET volumes $\hat{H}$ generated by $G$, the joint network enforces gate-to-gate consistency in the transformed synthetic HDPET for each target gate, providing additional supervision for training $G$. The details are as follows.

\noindent\textbf{Siamese Generative Network} is illustrated in Figure \ref{fig:network}. The Siamese generative network $G$ with encoding and decoding architecture is firstly supervised by a $\mathcal{L}_1$ loss, a structural similarity (SSIM) loss, and an adversarial loss to ensure the noise reduction and structure recovery. Specifically, we use a $\mathcal{L}_1$ loss to ensure the general appearance recovery and a $\mathcal{L}_{SSIM}$ loss to ensure the fine-detailed structure recovery. $\mathcal{L}_1$ loss allows noise suppression and SNR improvement, at the expense of reduced image sharpness. On the other hand, $\mathcal{L}_{SSIM}$ loss encourages image to have high contrast, sharpness and resolution. Given $L_{tgt}$ and $L_{ref}$ the target and reference LDPET gated volumes respectively, $G$ takes a pair of $[L_{tgt}, L_{ref}]$ and channel-wise concatenates each volumes with anatomical prior CT ($\rho$) to predict $\bar{H}_{tgt} = G (L_{tgt} , \rho ; \theta_G)$ and $\bar{H}_{ref} = G (L_{ref} , \rho ; \theta_G)$ simultaneously. The $\mathcal{L}_1$ loss and the $\mathcal{L}_{SSIM}$ loss can be written as:
\begin{equation} \footnotesize
    \mathcal{L}_1 = \sum_i || H_i - \bar{H}_i ||, \quad i \in \{ tgt,ref \}
\end{equation}
\begin{subequations}
\vspace{-0.1cm}
\begin{equation} \footnotesize
    \mathcal{L}_{SSIM} = \sum_i [1 - SSIM(H_i, \bar{H}_i)], \quad i \in \{ tgt,ref \} 
\end{equation}
\vspace{-0.1cm}
\begin{equation} \footnotesize
    SSIM(x,y) = \frac{2 m_x m_y + C_1}{m_x^2 + m_y^2 + C_1} \cdot \frac{2 \sigma_{xy} + C_2}{\sigma_x^2 + \sigma_y^2 + C_2}
\end{equation}
\end{subequations}
where $[m_x, m_y]$ and $[\sigma_x, \sigma_y]$ denote mean and standard deviation of an image pair $[x, y]$. The cross-covariance of $[x,y]$ is denoted as $\sigma_{xy}$. $C_1$ and $C_2$ are constant parameters. The adversarial loss from the discriminator $D$ provides an indication of discrepancy between prediction and ground truth as both $G$ and $D$ progressively optimized. Thus, the adversarial loss is also added to minimize the perceptual difference between prediction and ground truth from a CNN perspective. We utilize the adversarial loss in Wasseerstein GAN with gradient penalty (WGAN-GP) to achieve stable adversarial training \cite{arjovsky2017wasserstein}, which is formulated as:
\begin{align} \footnotesize
\begin{split}
    \mathcal{L}_{adv} = \sum_i \mathbb{E} [D(\bar{H}_i)] - \mathbb{E} [D(H_i)] + & \lambda_{gp} \mathbb{E} [(|| \nabla_{\ddot{H}_i} D(\ddot{H}_i) ||_2 - 1 )^2], i \in \{ tgt,ref \}
\end{split}
\end{align}
where $\ddot{H}$ represents a linear combination of $\bar{H}$ and $H$ with a weight $t$ uniformly sampled between $0$ and $1$. Thereby, $\lambda_{gp}$ controls the gradient penalty level and is set to $3$ here. The combination of these three loss functions formulates our Structure Recovery (SR) loss as:
\begin{equation} \footnotesize
    \mathcal{L}_{SR} = \beta_1 \mathcal{L}_1 + \beta_2 \mathcal{L}_{SSIM} + \beta_3 \mathcal{L}_{adv}
\end{equation}
where $\beta_1$, $\beta_2$, and $\beta_3$ are loss weights. In our experiments, we empirically set $\beta_1 = 1$, $\beta_2 = 1$, and $\beta_3 = 0.2$ for balance training. 

\noindent\textbf{Motion estimation network}
$R$ aims to predict the transformation between target and reference gated volumes. Here, we use a probabilistic generative model \cite{dalca2018unsupervised} to predict the transformation, as illustrated in Figure \ref{fig:network}'s left section. Assuming $H_{ref}$ and $H_{tgt}$ are volumes that need to be registered and the transformation between them is parameterized by a sampled velocity field $V$, $R$ aims to find the most likely registration field by optimizing the posterior probability $p(V | H_{ref}, H_{tgt})$. Thus, the loss function for network $R$ can be derived and written as:
\begin{align} \footnotesize
\begin{split}
      \mathcal{L}_R (H_{ref},H_{tgt}) = & \frac{1}{K} \sum_k ||H_{ref} - T \circ H_{tgt}|| + \text{KL} [q_{\theta_R} (V|H_{ref},H_{tgt}) || p(V)]
\end{split}
\end{align}
where $K$ is the number of samples in each training batch, $T$ is the transformation function parameterized by $V \sim q_{\theta_R} (V|H_{ref},H_{tgt})$. The first term minimizes the L1 distance between reference volume $H_{ref}$ and warped target volume $H_{tgt}$. The second term ensures the distribution similarity between posterior and prior of $V$. $\mathcal{L}_R$ is the transform consistency loss in Figure \ref{fig:network}. During the inference stage, the predicted $V$ is fed into the scaling and squaring layer \cite{arsigny2006log} to integrate $V$ over $[0,1]$, and produce the final transformation $T$. Then, $T$ and the target volume $H_{tgt}$ are inputted into the spatial transform layer to generate the warped target volume $T \circ H_{tgt}$. Detailed derivation is in our supplementary. 

\noindent\textbf{Gate-to-Gate Consistency Learning}
The Siamese generator in the first part maps $L$ to $H$ with SR loss $\mathcal{L}_{SR}$ for individual gates. However, the appearance consistency constraint between gates is not utilized. A gate-to-gate consistency should sustain when the gated volumes are registered. The gate-to-gate consistency learning is achieved by feeding the synthetic pair of HDPET volumes, $[\bar{H}_{tgt}, \bar{H}_{ref}]$ generated using the Siamese generative network $G$, into the pre-trained motion estimation network $R$ after concatenating these two networks. Therefore, the transformation prediction process of the joint network can be described as:
\begin{equation} \footnotesize
    \bar{T} = R(\bar{H}_{ref}, \bar{H}_{tgt} ; \theta_R) = R(G (L_{ref} , \rho ; \theta_G), G (L_{tgt} , \rho ; \theta_G) ; \theta_R)
\end{equation}
Given the transformation $\bar{T}$, we warp the synthetic $\bar{H}_{tgt}$ and obtain $\bar{T} \circ \bar{H}_{tgt}$. We aim to minimize the distance between $\bar{T} \circ \bar{H}_{tgt}$ and ground truth $H_{ref}$, such that the transformed target gated volume and reference gated volume are consistent. Thus, the gate-to-gate transform consistency loss can be formulated as:
\begin{align} \footnotesize
\begin{split}
      \mathcal{L}_{G2G} =  \frac{1}{K} \sum_k ||H_{ref} - \bar{T} \circ \bar{H}_{tgt}|| + \text{KL} [q_{\theta_R} (V|\bar{H}_{ref},\bar{H}_{tgt}) || p(V)]
\end{split}
\end{align}
The first term encourages the gate-to-gate appearance consistency using a $\mathcal{L}_1$ norm and the second term ensures the distribution similarity between posterior and prior of $V$. $\mathcal{L}_{G2G}$ provides additional supervision for optimizing $G$ by utilizing the inter-gate relationship. It is the key in our Siamese network design that enables us to randomly sample pairs of gated volume, which augments the number of available training data for each subject to $A_6^2 = 30$. Therefore, the denoising and structural recovery from LDPET to HDPET will be more reliable.

Finally, our full loss function for optimizing $G$ is $\mathcal{L}_{tot} = \mathcal{L}_{SR} + \mathcal{L}_{G2G}$, which is trained in an adversarial manner. $G$ and $R$ try to minimize this loss collaboratively, while $D$ tries to maximize it. To optimize the overall network, we update $G$, $R$, and $D$ alternatively by: optimizing $D$ with $G$ and $R$ fixed, then optimizing $G$ with $D$ and $R$ fixed.

\subsection{Evaluation with Human Data}
We collected 29 pancreas \textsuperscript{18}F-FPDTBZ \cite{normandin2012vivo} PET/CT studies with respiration gating facilitated by the Anzai system. The total acquisition time was $120$ mins for each study. We used phase gating to generate 6 gates for each study. To eliminate the mismatch between attenuation correction (AC) map and gated PET, instead of using CT as AC-map, we utilized the maximum likelihood estimation of activity and attenuation (MLAA) \cite{rezaei2016simultaneous} to generated AC-map for each gated volume to ensure phase-matched attenuation correction, where CT was used as initial estimation for MLAA iterations. The HDPET volumes were reconstructed with 100\% of the listmode data mimicking high radiation dose data. The LDPET volumes were reconstructed with 1.5\% of the listmode data with random sampling. Each data was reconstructed into a $400 \times 400 \times 109$ volume with voxel size of $2.032 \times 2.032 \times 2.027$ $mm^3$. The central $200 \times 200 \times 109$ voxels were kept to remove most voxels outside the human body contour and resized to $128 \times 128 \times 128$. The end expiration gate (typically Gate 4) was used as the reference gate since it shows minimum intra-gate motion. 

The dataset were split into training set of 22 studies and test set of 7 studies. The evaluation was performed on the 7 test studies with 6 gated volumes in each study. For quantitative evaluation, the denoising results were evaluated by comparing the synthetic HDPET volumes to the ground truth HDPET volumes using the Peak Signal-to-Noise Ratio (PSNR) and Structural Similarity Index (SSIM). The motion estimation results were evaluated using the Mean Vector Euclidean Distance (MVED) that measures the 3D Euclidean distance between the predicted vector field and the ground truth vector field, which was defined as the motion field predicted by $R$ using the ground truth HDPET. For comparative study, we compared our results against the following algorithms: Gaussian filtering (GAU), Non-local mean filtering (NLM) \cite{dutta2013non}, Block-matching 4D filtering (BM4D) \cite{maggioni2012nonlocal}, UNet \cite{lu2019investigation,ronneberger2015u}, and cGAN \cite{wang20183d}. 

\begin{figure}[htb!]
\centering
\includegraphics[width=1.00\textwidth]{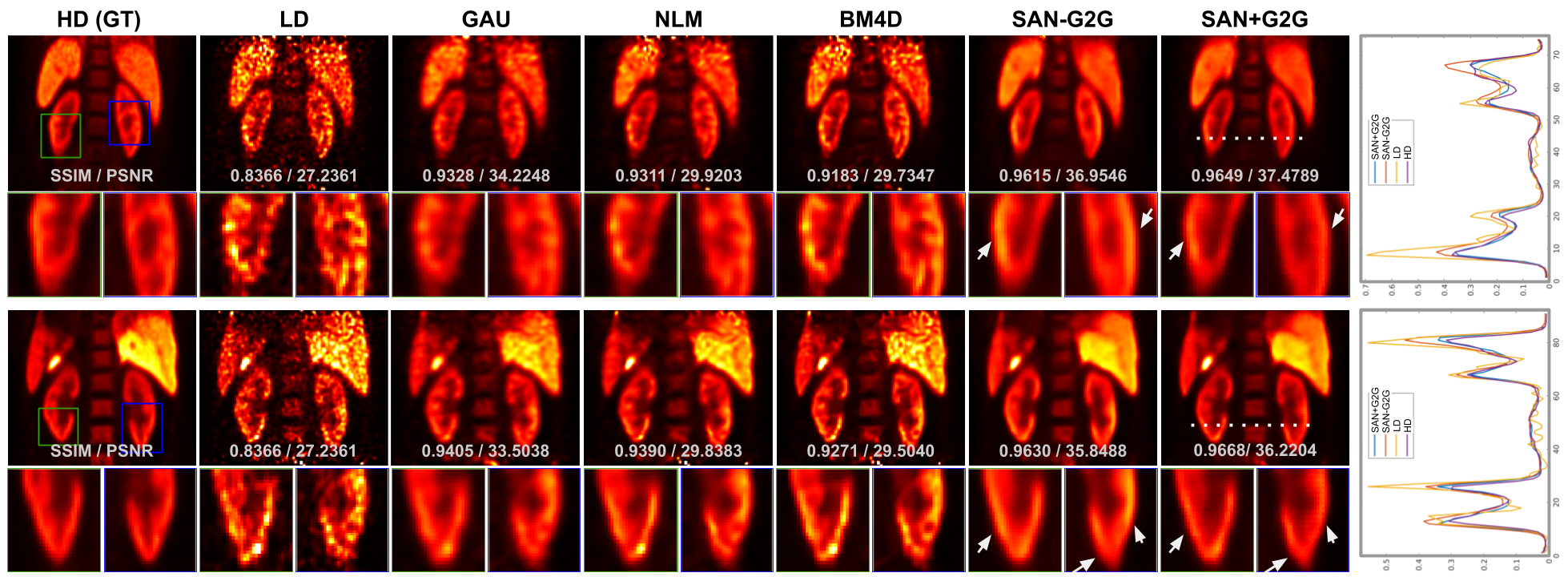}
\caption{Sample HD and 1.5\% LD PET slices with enlarged subregions using various denoising methods for Two sample subjects. The corresponding PSNR and SSIM are indicated at the bottom of the images. Comparison of intensity profile is also shown on the right. -/+G2G denotes without/with gate-to-gate consistency learning.}
\label{fig:comp_dn}
\end{figure}

\section{Results}
The qualitative comparison of various denoising methods is shown in Figure \ref{fig:comp_dn}. As we can observe on the figure, conventional post-processing methods, such as NLM \cite{dutta2013non} and BM4D \cite{maggioni2012nonlocal}, have difficulties in structural recovery when only 1.5\% of the total counts was considered. The high noise level also introduced additional artifacts, resulting in inferior performance compared to the standard Gaussian filtering. In contrast, deep learning based methods achieved better performance in noise reduction and structural recovery. 

\begin{figure}
\centering
\begin{minipage}{.49\textwidth}
  \centering
  \includegraphics[width=1.00\linewidth]{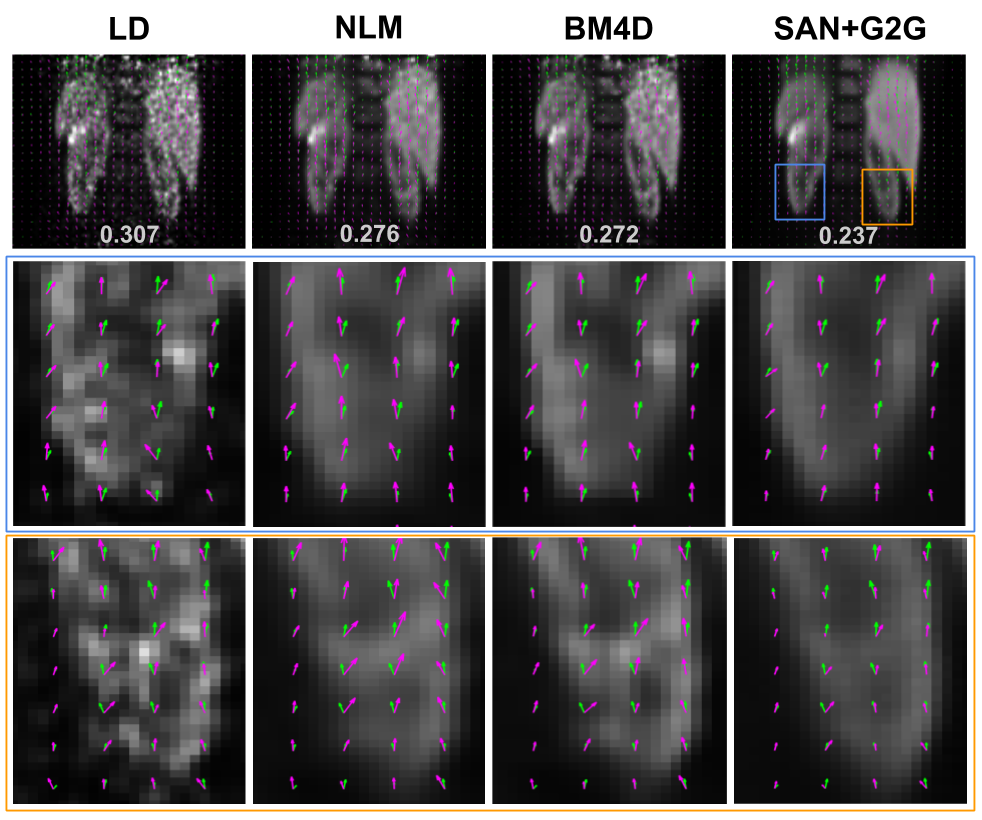}
  \captionof{figure}{Qualitative motion estimation results from different denoising methods. Ground truth (green arrows) and predicted (magenta arrows) motion estimation vectors are overlaid on denoised images.}
  \label{fig:motion}
\end{minipage}
\begin{minipage}{.49\textwidth}
  \centering
  \includegraphics[width=1.00\linewidth]{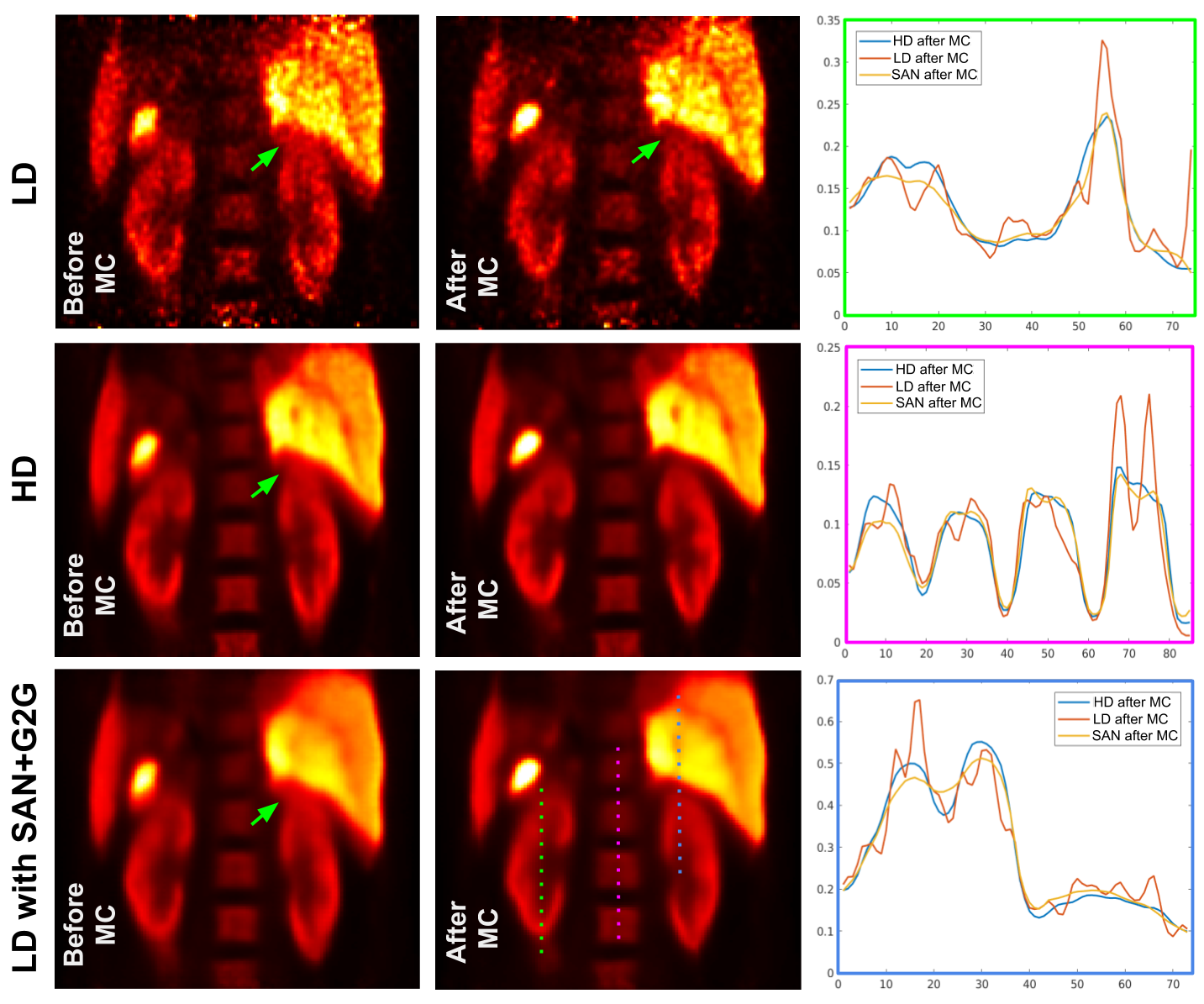}
  \captionof{figure}{Illustration of motion blurred images (left) and averaged image of all gates average registered to the reference frame (middle). The green arrows indicate where significant motion reduction is observed after applying the proposed SAN+G2G.}
  \label{fig:final_average}
\end{minipage}
\end{figure}

Table \ref{tab:PSNRandSSIM} outlines the quantitative comparison of different methods on PET image denoising. Both PSNR and SSIM were evaluated for each gated volumes ($\mathcal{G}_n$), along with averaged value computed on the last column. Among them, our SAN without G2G outperforms the previous deep learning based methods, and the addition of G2G learning that utilizes the information over gates further improved the performance. In parallel, Figure \ref{fig:motion} illustrates a qualitative comparison of motion estimation based on the discussed denoising methods. As we can see, our proposed SAN+G2G yields the most consistent motion vectors between the estimated and ground truth motion vectors. The quantitative comparison of motion estimation among different denoising methods is given in Table \ref{tab:motion}. As shown in the table, our SAN+G2G was able to improve the motion estimation accuracy by $20\%$ in average, achieving the lowest $0.264$ in averaged MVED, compared to other studied methods. Using our proposed method, denoised gated LDPET volumes can be generated with corresponding motion vectors to the reference gate. We then registered all gated volumes of LDPET, HDPET, and LDPET with SAN+G2G to the reference gate.  As an example shown in Figure \ref{fig:final_average}, the proposed network is able to generate gated PET volumes with reduced noise level and a final motion corrected image that averaged all registered image volumes with reduced motion blurring using low dose gated data.

\begin{table}
\scriptsize
\centering
\caption{Quantitative comparison of denoising results using PSNR (dB) and SSIM ($\times 10^2$). Among conventional post-processing methods and deep learning based methods, the optimal results are marked in \textcolor{red}{red}.}
\label{tab:PSNRandSSIM}
    \begin{tabular}{|c||c|c|c|c|c|c||c|}
        \hline 
        \textbf{PSNR/SSIM} & $\mathcal{G}1$ & $\mathcal{G}2$ & $\mathcal{G}3$ & $\mathcal{G}4$ & $\mathcal{G}5$ & $\mathcal{G}6$ & Average \Tstrut\Bstrut\\
        \hline
        LDPET              & 28.48/86.3 & 28.47/86.5 & 28.27/86.6 & 28.47/86.9 & 27.93/86.1 & 28.11/86.2 & 28.29/86.4 \Tstrut\Bstrut\\
        \hline
        GAU                & 34.92/94.6 & 34.25/94.5 & 34.18/94.5 & 34.32/94.6 & 34.36/94.4 & 34.66/94.6 & 34.45/94.6 \Tstrut\Bstrut\\
        NLM \cite{dutta2013non}   & 31.71/94.5 & 31.80/94.5 & 31.39/94.4 & 31.60/94.6 & 31.03/94.3 & 31.23/94.5 & 31.46/94.5 \Tstrut\Bstrut\\
        BM4D \cite{maggioni2012nonlocal}  & 31.32/93.5 & 31.32/93.5 & 31.03/93.5 & 31.21/93.8 & 30.68/93.4 & 30.83/93.4 & 31.07/93.5 \Tstrut\Bstrut\\
        \hline
        UNet \cite{lu2019investigation}   & 37.12/95.9 & 36.01/95.8 & 36.24/95.7 & 36.32/96.1 & 36.34/95.9 & 36.85/96.1 & 36.48/95.9 \Tstrut\Bstrut\\
        cGAN \cite{wang20183d}       & 37.38/96.3 & 36.21/96.2 & 36.41/96.1 & 36.43/96.1 & 36.58/96.2 & 37.02/96.2 & 36.67/96.2 \Tstrut\Bstrut\\
        \hline
        SAN-G2G           & 37.55/96.8 & 36.48/96.6 & 36.59/96.6 & 36.67/96.7 & 36.96/96.6 & 37.25/96.8 & 36.92/96.7 \Tstrut\Bstrut\\
        SAN+G2G           & \textcolor{red}{37.81/97.1} & \textcolor{red}{36.74/96.9} & \textcolor{red}{36.77/96.9} & \textcolor{red}{36.87/96.9} & \textcolor{red}{37.35/97.0} & \textcolor{red}{37.43/97.1} & \textcolor{red}{\textbf{37.16/97.0}} \Tstrut\Bstrut\\
        \hline
    \end{tabular}
\end{table}

\begin{table}[htb!]
\scriptsize
\centering
\caption{Quantitative comparison of motion estimation results evaluated in terms of MVED. G4 is the reference gate. Optimal results are marked in \textcolor{red}{red}.}
\label{tab:motion}
    \begin{tabular}{|c||c|c|c|c|c|c||c|}
        \hline
        \textbf{MVED} & $\mathcal{G}1$ & $\mathcal{G}2$ & $\mathcal{G}3$ & $\mathcal{G}4$ (Ref) & $\mathcal{G}5$ & $\mathcal{G}6$ & Average \Tstrut\Bstrut\\
        \hline
        LDPET         & 0.392 & 0.342 & 0.289 & - & 0.282 & 0.350 & 0.331 \Tstrut\Bstrut\\
        GAU           & 0.368 & 0.372 & 0.328 & - & 0.328 & 0.352 & 0.349 \Tstrut\Bstrut\\
        NLM \cite{dutta2013non}   & 0.360 & 0.363 & 0.312 & - & 0.309 & 0.342 & 0.337 \Tstrut\Bstrut\\
        BM4D \cite{maggioni2012nonlocal}  & 0.351 & 0.341 & 0.2997 & - & 0.287 & 0.331 & 0.322 \Tstrut\Bstrut\\
        SAN-G2G       & 0.309 & 0.309 & 0.262 & - &  0.263 & 0.297 & 0.288 \Tstrut\Bstrut\\
        SAN+G2G       & \textcolor{red}{0.289} & \textcolor{red}{0.285} & \textcolor{red}{0.236} & - & \textcolor{red}{0.237} & \textcolor{red}{0.274} & \textcolor{red}{\textbf{0.264}} \Tstrut\Bstrut\\
        \hline
    \end{tabular}
\end{table}

\section{Discussion and Conclusion}
In this work, we propose a Siamese adversarial network with gate-to-gate consistency learning, a novel framework for low dose gated PET denoising and motion estimation, simultaneously. We first pre-train our motion estimation network on the ground truth HDPET, and concatenate it to our Siamese adversarial network that enables the gate-to-gate consistency learning for improving the denoising performance. The denoised low-dose gated volumes are then fed into the motion estimation network for robust motion estimation. In our framework, the Siamese input design allows us to efficiently augment the training data from each patient, thus can better train generalizable denoising and motion estimation models. We demonstrated the feasibility of our method on the tasks of PET image denoising and motion estimation with promising performance. 

The potential clinical feasibility of our work is two-fold. Firstly, as high-noise level and motion are inevitable in the chest and abdominal low-dose PET acquisitions, it will affect the visualization of small pathological findings, such as lung/liver lesions. Our work is potentially useful for recovering these small objects from noise and correcting motions to improve the delineation of distorted objects. Secondly, the estimated motion can be incorporated into the motion compensated PET reconstruction frameworks toward motion-free low-dose PET reconstructions, which will also improve the reconstruction quality by reducing the motion artifacts. We will explore these directions in our future works.

\bibliographystyle{splncs}
\bibliography{bibliography}

\end{document}


\title{Supplemental Materials}
 

\author{Bo Zhou\inst{1} \and Yu-Jung Tsai\inst{2} \and Chi Liu\inst{1,2}}

\institute{Biomedical Engineering, Yale University, New Haven, CT, USA \and Radiology and Biomedical Imaging, Yale University, New Haven, CT, USA}

\authorrunning{B. Zhou, etc} 
\titlerunning{Denoising and motion estimation for low-dose gated PET} 

\maketitle              




\section{Network Architecture \& Implementation Details}
G, R, and D's network architectures are summarized in Table \ref{tab:network_arch}. Both G and R use a UNet backbone structure. 

To avoid overfitting, we deployed two augmentation techniques: 1) we applied identical random cropping and 90 degrees rotation along x,y,z axis for the Siamese input, and 2) we randomly chose 2 gates from 6 gates of each patient for Siamese input, which resulted in $A^2_6 = 30$ training pairs with each patient data. The Adam solver was used to optimize the loss functions in this work with a momentum of 0.99 and learning rate of 0.0001. The network was trained on a Quadro RTX 8000 GPU with 48GB memory.

\begin{table} [htb!]
\centering
\caption{Configuration details of generator (G), discriminator (D), and motion estimator (R) in our Siamese Adversarial Network. The input size is denoted as (batchsize $\times$ width $\times$ height $\times$ depth $\times$ channel). The operations are denoted as LReLU: Leaky-ReLU; BN: Batch Normalization; Concat: Concatenation along channel axis; Upsample2: $\times$2 upsampling; FC: fully connected layer. Convolutional layer and deconvolutional layer are presented in the form of Conv3D(in-channel, out-channel, kernel size, stride, padding) and DeConv3D(in-channel, out-channel, kernel size, stride). $|$ denotes skip connection.}
\label{tab:network_arch}
\resizebox{\textwidth}{!}{
    \begin{tabular}{l  l  l}
        \hline
        \textbf{G}                                        & \textbf{D}                         & \textbf{R}                \\ [0.025cm]
        \hline
        Inputs: (nx128x128x128x1)                                   & Inputs: (nx128x128x128x1)          & Inputs: (nx128x128x128x1)                                                 \\ [0.01cm]
        LReLU(BN(Conv3D(1,8,3,1,1)))                                & LReLU(Conv3D(1,16,3,2,1))          & LReLU(Conv3D(1,16,3,2,1))                                                 \\ [0.005cm]
        $| \quad$ LReLU(BN(Conv3D(8,16,3,2,1)))                     & LReLU(BN(Conv3D(16,32,3,2,1)))     & $| \quad $ LReLU(Conv3D(16,32,3,2,1))                                     \\ [0.005cm]
        $| \quad | \quad$ LReLU(BN(Conv3D(16,32,3,2,1)))            & LReLU(BN(Conv3D(32,64,3,2,1)))     & $| \quad | \quad $ LReLU(Conv3D(32,32,3,2,1))                             \\ [0.005cm]
        $| \quad | \quad | \quad$ LReLU(BN(Conv3D(32,32,3,2,1)))    & LReLU(BN(Conv3D(64,128,3,2,1)))    & $| \quad | \quad | \quad $ LReLU(Conv3D(32,32,3,2,1))                     \\ [0.005cm]
        $| \quad | \quad | \quad$ ReLU(BN(Conv3D(32,64,3,1,1)))     & LReLU(BN(Conv3D(128,256,3,2,1)))   & $| \quad | \quad | \quad | \quad $ Upsample2(LReLU(Conv3D(32,32,3,1,1)))  \\ [0.005cm]
        $| \quad | \quad | \quad$ DeConv3D(64,64,2,2,0))            & FC(Flatten())                      & $| \quad | \quad | \quad$ Concat()                                        \\ [0.005cm]
        $| \quad | \quad$ Concat()                                  &                                    & $| \quad | \quad | \quad$ Upsample2(LReLU(Conv3D(32+32,32,3,1,1)))        \\ [0.005cm]
        $| \quad | \quad$ ReLU(BN(Conv3D(64+32,64,3,1,1)))          &                                    & $| \quad | \quad$ Concat()                                                \\ [0.005cm]
        $| \quad | \quad$ DeConv3D(64,32,2,2,0))                    &                                    & $| \quad | \quad$ Upsample2(LReLU(Conv3D(32+32,32,3,1,1)))                \\ [0.005cm]
        $| \quad$ Concat()                                          &                                    & $| \quad $ Concat()                                                       \\ [0.005cm]
        $| \quad$ ReLU(BN(Conv3D(32+16,32,3,1,1)))                  &                                    & $| \quad $ LReLU(Conv3D(32+32,16,3,1,1))                                  \\ [0.005cm]
        $| \quad$ DeConv3D(32,16,2,2,0))                            &                                    & Concat()                                                                  \\ [0.005cm]
        Concat()                                                    &                                    & Conv3D(16+16,1,1,1,0)                                                     \\ [0.005cm]
        ReLU(BN(Conv3D(16+8,16,3,1,1)))                             &                                    &                                                                           \\ [0.005cm]
        ReLU(BN(Conv3D(16,8,3,1,1)))                                &                                    &                                                                           \\ [0.005cm]
        ReLU(BN(Conv3D(8,1,1,1,0)))                                 &                                    &                                                                           \\ [0.005cm]
        \hline
    \end{tabular}
    }
\end{table}

\section{Motion Estimation Network}
Denoting $H_{ref}$ and $H_{tgt}$ as two volumes need to be registered and $V$ a sampled stationary velocity field that parameterizes a transformation, the goal is to compute the posterior probability $p(V|H_{ref},H_{tgt})$ such that we can get the most likely registration field for a volume pair $[H_{ref},H_{tgt}]$. $R$ generates $\mu_{V|H_{ref},H_{tgt}}$ and $\sigma_{V|H_{ref},H_{tgt}}$ for sampling $V$ that specifies a diffeomorphism. Assuming the prior probability $p(V)$ and modeled posterior $q_{\theta_R}$ of $V$ are both multivariate normal distributions, we have:
\begin{equation}
    p(V)=\mathcal{N}(V;0,\sigma_V)
\end{equation}
\begin{equation}
    q_{\theta_R}=\mathcal{N}(V;\mu_{V|H_{ref},H_{tgt}},\sigma_{V|H_{ref},H_{tgt}})
\end{equation}
Thus, the KL divergence can be computed as:
\begin{align} \label{eq:kl}
\begin{split}
    \min_{\theta_R} & KL [q_{\theta_R}(V|H_{ref},H_{tgt}) || p(V|H_{ref},H_{tgt})] \\
                & = \min_{\theta_R} \mathbb{E}_q [\log q_{\theta_R} (V|H_{ref},H_{tgt}) - \log p (V|H_{ref},H_{tgt})] \\
                & = \min_{\theta_R} \mathbb{E}_q [\log q_{\theta_R} (V|H_{ref},H_{tgt}) - \log \frac{p(V,H_{ref},H_{tgt})}{p(H_{ref},H_{tgt})}] \\
                & = \min_{\theta_R} \mathbb{E}_q [\log q_{\theta_R} (V|H_{ref},H_{tgt}) - \log p(V)] - \mathbb{E}_q [\log p(H_{ref}|V,H_{tgt})] \\
                & = \min_{\theta_R} KL [q_{\theta_R} (V|H_{ref},H_{tgt}) || p(V)] - \mathbb{E}_q [\log p(H_{ref}|V,H_{tgt})]
\end{split}
\end{align}
Then, we can train $R$ by optimizing the variational lower bound from the above equation. Thereby, the loss function can be re-written as:
\begin{align}
\begin{split}
      \mathcal{L}_R (H_{ref},H_{tgt}) = & - \mathbb{E}_q [\log p(H_{ref}|V,H_{tgt})] + KL [q_{\theta_R} (V|H_{ref},H_{tgt}) || p(V)] \\
                    = & \frac{1}{K} \sum_k ||H_{ref} - T \circ H_{tgt}|| + KL [q_{\theta_R} (V|H_{ref},H_{tgt}) || p(V)]
\end{split}
\end{align}
where $K$ is the number of sample in each training batch. $T$ is the transformation function parameterized by $V \sim q_{\theta_R} (V|H_{ref},H_{tgt}) = \mathcal{N}(V;\mu_{V|H_{ref},H_{tgt}},\sigma_{V|H_{ref},H_{tgt}})$. The first term minimizes the L1 distance between the reference volume $H_{ref}$ and the warped target volume $T \circ H_{tgt}$. The second term ensures the distribution similarity between posterior and prior of $V$. 

\section{Additional Results}
\begin{figure}
  \centering
  \includegraphics[width=1.00\linewidth]{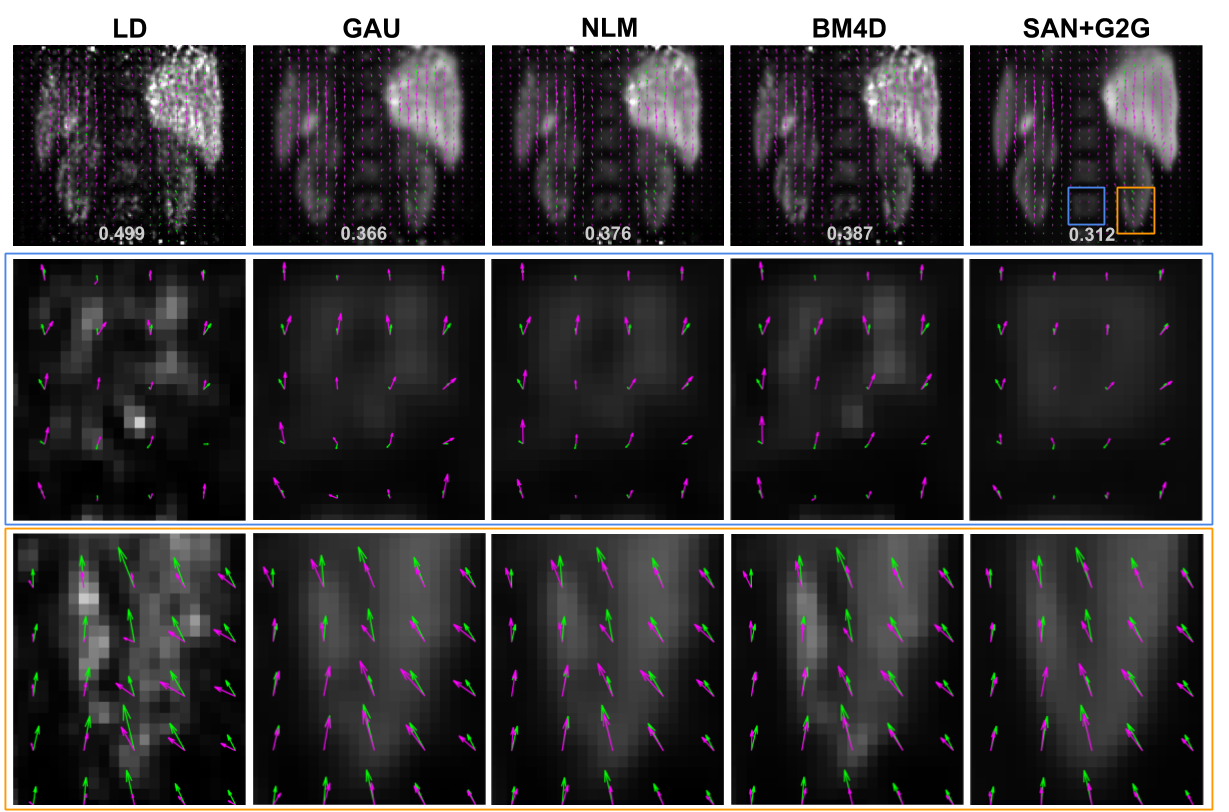}
  \captionof{figure}{Qualitative illustration of motion estimation results from different denoising methods. Ground truth (green arrows) and predicted (magenta arrows) motion estimation vectors are overlaid on the denoised images.}
  \label{fig:motion1}
\end{figure}

\begin{figure}
  \centering
  \includegraphics[width=1.00\linewidth]{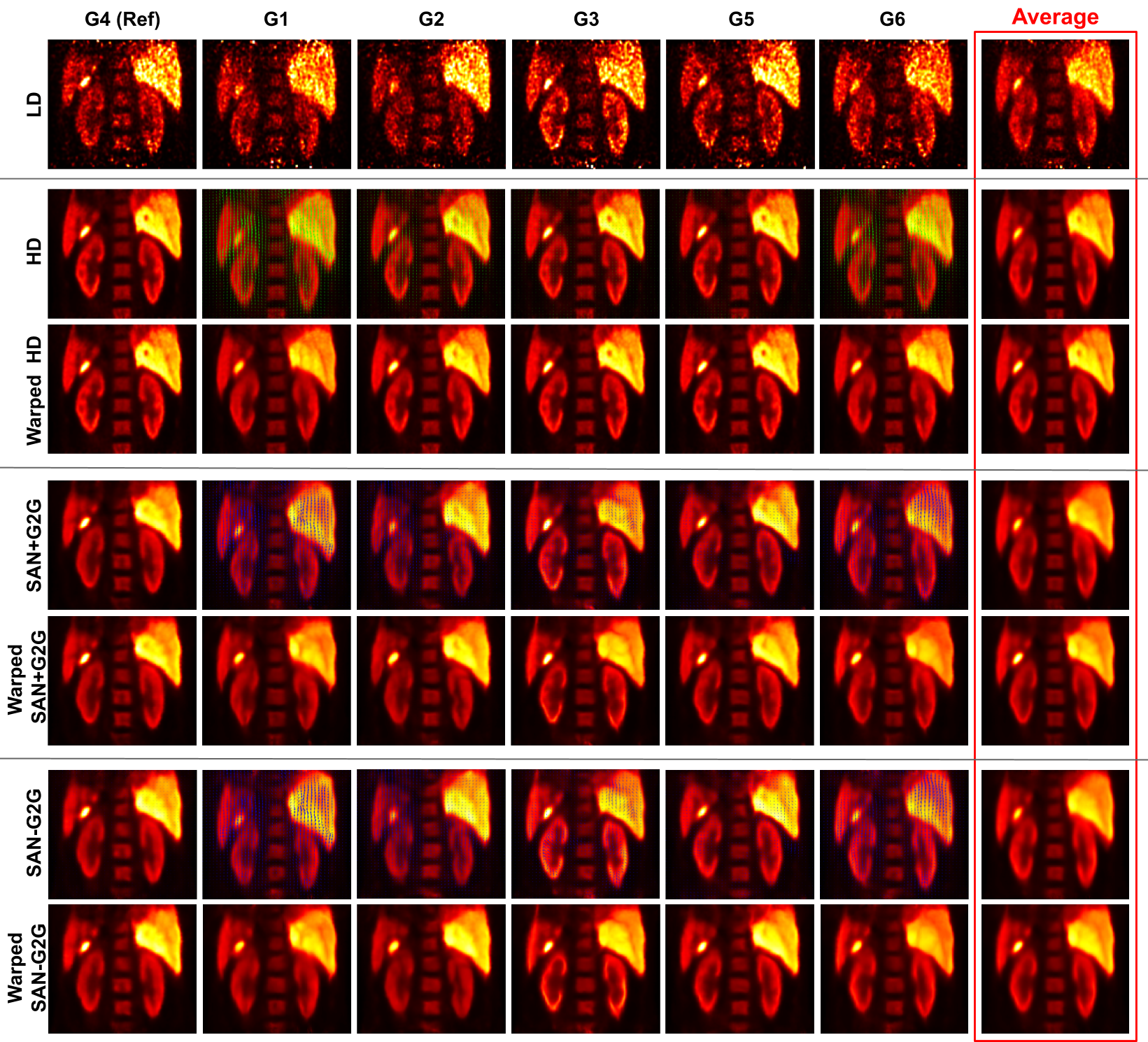}
  \captionof{figure}{Illustration of each gates' denoising and motion estimation results. The average images of all gates with and without applying the corresponding motion transformation that deforms each gate to align it with the reference gate are shown on the right. Gate 4 (G4) is used as reference gate here. LD: low dose volume, HD: high dose volume.}
  \label{fig:final_average1}
\end{figure}

Additional motion estimation results are shown in Figure \ref{fig:motion1}. As we can see from the comparison, our SAN+G2G can produce motion estimation much closer to the ground truth motion. Additional denoising result for each gate as well as the corresponding motion estimation results are shown in Figure \ref{fig:final_average1}. The average images of all gates with and without applying the corresponding transformation $T$ that deforms each gate to align it with the reference gate are also provided.

\bibliographystyle{splncs}
\bibliography{bibliography}